\documentclass[aps,prb,twocolumn,showpacs,amsmath,amssymb]{revtex4-1}
\usepackage{graphicx}
\usepackage{bm}
\usepackage{amsmath}
\usepackage{color}
\usepackage[caption=false]{subfig}
\usepackage{url}
\usepackage[pdftex,colorlinks=true,linkcolor=blue,citecolor=blue,urlcolor=blue,breaklinks]{
hyperref}

\newcommand{\be}{\begin{equation}}
\newcommand{\ee}{\end{equation}}
\newcommand{\bea}{\begin{eqnarray}}
\newcommand{\eea}{\end{eqnarray}}
\newcommand{\up}{\uparrow}
\newcommand{\down}{\downarrow}
\newcommand{\bwt}{\begin{widetext}}
\newcommand{\ewt}{\end{widetext}}
\newcommand{\ham}{\mathcal{H}}

\newcommand{\gn}{\mathcal{G}}
\newcommand{\ra}{\rangle}
\newcommand{\la}{\langle}
\newcommand{\bsb}{\begin{subarray}}
\newcommand{\esb}{\end{subarray}}
\newcommand{\largem}{\!\!}

\newcommand{\eins}{\mbox{$1 \hspace{-1.0mm} {\bf l}$}}
\newcommand{\iu}{{i\mkern1mu}}

\newcommand{\vecv}[2]{
\left(\largem
 \begin{tabular}{c}
  $#1$ \\
  $#2$
  \end{tabular}
  \largem
\right)
}

\newcommand{\vech}[2]{
\left(\largem
 \begin{tabular}{c}
  $#1$ \! $#2$
  \end{tabular}
  \largem
\right)
}

\newcommand{\mat}[4]{
\left(
\largem
 \begin{tabular}{cc}
  $#1$ & $#2$ \\
  $#3$ & $#4$
  \end{tabular}
  \largem
\right)
}

\begin{document}

\title{Disordered graphene Josephson junctions}

\author{W. A. Mu\~noz}
\email{WilliamArmando.Munoz@uantwerp.be}
\author{L. Covaci}
\email{lucian@covaci.org}
\author{F. M. Peeters}
\email{Francois.Peeters@uantwerp.be}
\affiliation{Departement Fysica, Universiteit Antwerpen, Groenenborgerlaan 171,
B-2020 Antwerpen, Belgium}

\date{\today}

\begin{abstract}
A tight-binding approach based
on the Chebyshev-Bogoliubov-de Gennes method is used to describe disordered single-layer graphene
Josephson junctions. Scattering by vacancies, ripples or charged impurities is included. We compute
the Josephson current and investigate the nature of multiple Andreev reflections, which induce bound
states appearing as peaks in the density of states for energies
below the superconducting gap. In the presence of single atom vacancies, we observe a strong
suppression of the supercurrent that is a consequence of strong inter-valley scattering. Although
lattice deformations should not induce inter-valley scattering, we find that the supercurrent is
still suppressed, which is due to the presence of pseudo-magnetic barriers. For charged impurities,
we consider two cases depending on
whether the average doping is zero, i.e. existence of electron-hole puddles, or finite. In both
cases, short range impurities strongly affect the supercurrent, similar to the vacancies
scenario.
\end{abstract}



\maketitle

%
\section{Introduction}
The notable absence of intrinsic superconductivity in graphene
has not been an obstacle for recent experimental advances demonstrating
potential
applications of graphene in superconducting devices by using the proximity
effect
~\cite{heersche_bipolar_2007,komatsu_superconducting_2012,
Mizuno2013,Choi2013, deon_tuning_2014}.
Despite the fact that the interplay between superconductivity and quantum
relativistic dynamics
in graphene, expressed in an unusual Andreev reflection, has been elusive to
experiments,
advances in lithography make graphene-based superconducting devices a possible
platform for
superconducting quantum engineering. However, it has been observed that
superconducting states
in graphene are strongly affected by the inherent disorder that is present in graphene
samples. More
relevant,  the specular Andreev reflection predicted to take place in a clean
superconducting-normal (S/N) graphene interface and where, different from the
conventional
retro-reflection, the path of the reflected hole does not retrace the path of
the incoming electron,
cannot be observed in the presence of dopant inhomogeneities.

Many speculations have been made on the effect of
disorder, like for instance the report of a gate-tunable Josephson junction
where the off state at the Dirac point is believed to be caused by the suppression of the
supercurrent
due to intrinsic ripples appearing in graphene~\cite{Choi2013}. A
suppression
of the critical current due to the presence of puddles of charges has been reported
as
well~\cite{komatsu_superconducting_2012}. Thus, disorder can
play an important role
in graphene superconducting devices.

From a theoretical point of view, the interplay between superconductivity and
disorder in graphene
has not been thoroughly investigated. In only a few exceptional cases, works showing
the role of disorder in intrinsic
superconductivity~\cite{nandkishore_superconductivity_2013}, as
well as in S/N graphene
interfaces~\cite{cheng_effect_2011,burset_proximity-induced_2009} have been
recently reported. For instance, a counter-intuitive enhancement of
superconductivity by weak
disorder has been predicted, while others have shown that the presence of
disorder
prevents the observation of the specular Andreev reflection and suppresses
intrinsic
superconductivity in graphene.
From the perspective of the continuum Dirac approximation only scattering
processes which mix the
$K$ and $K'$ valleys, are predicted to matter \cite{beenakker_colloquium:_2008}.
However, it has
been shown that inhomogeneous strain, which breaks the effective time-reversal
symmetry in each
cone but not the true time-reversal symmetry, can lead to the suppression of the
Cooper diffusion in
a graphene Josephson junction~\cite{covaci_superconducting_2011} by generating a
pseudo-magnetic
barrier and allowing suppercurrents to flow only as edge states.

The effect of disorder in graphene has been widely
studied~\citep{castro_neto_electronic_2009,mucciolo_disorder_2010} and was shown
to break various
symmetries, like the chiral or the effective time-reversal symmetries.
The absence of these symmetries may strongly affect electronic
transport~\cite{peres_electronic_2006,mucciolo_disorder_2010,Morpurgo2006}.
The first and most commonly investigated type of extrinsic disorder corresponds
to charge
inhomogeneities~\cite{ando_screening_2006,peres_electronic_2006}. This type of
disorder resembles
charged puddles, which are usually present when graphene is put on a
substrate~\cite{chen_charged-impurity_2008}, or when e.g. water
molecules are deposited on its surface. Depending on the strength of the potential
generated by these
charged impurities, or the distance from the graphene sheet, they can be considered
as long-range or
short-range potentials. For instance, elastic scattering from a short-range disorder potential may
mix
electron states in
different valleys $K$ and $K'$, i.e. inter-valley scattering. Instead, for a
long-range potential
varying smoothly over scales larger than the lattice constant, electrons in the $K$ and $K'$
valleys do
not mix.

Scattering on vacancies, i.e. the absence of a carbon
atom~\cite{pereira_disorder_2006,peres_electronic_2006,pereira_modeling_2008}, unlike charged
potentials, induces resonant states near the Dirac point.
This short-range
unitary scatterer may introduce a localized state, for which the wavefunction is
formed equally from
both $K$ and $K'$~\cite{pereira_disorder_2006} valleys, similar to the nature of
the wave-function
at armchair edges. For this type of disorder, coupling between valleys is
expected to
ocurr~\cite{park_formation_2011} and thus have a strong influence on the
supercurrent.

Yet another type of disorder is induced by lattice distortions, either intrinsic
or designed by
strain engineering. Due to its exceptional flexibility graphene can be easily
deformed by mechanical
stress or conform to the geometry of the substrate. Unusual high pseudo-magnetic
fields have been predicted to emerge from strained
graphene~\cite{levy_strain-induced_2010}. In
fact, theoretical descriptions have revealed the existence of an effective
vector potential coming
from the change in the hopping parameter due to the geometrical deformation of
the distance
between nearest neighbors carbon
atoms~\cite{de_juan_charge_2007,guinea_energy_2009,pereira_strain_2009,
guinea_generating_2010,
moldovan_electronic_2013}. Particularly, some works have investigated the
interplay between
superconductivity and uniform strain~\cite{alidoust_tunable_2011} or
pseudo-quantum Hall states in
graphene Josephson junctions~\cite{covaci_superconducting_2011}.

Both electrons and holes experience normal scattering inside the junction,
therefore different
dephasing mechanisms are expected to strongly influence the transmission of the
Copper pair between
the superconducting leads. Since the more general description reported so far is
based on the
continuum Dirac approximation~\cite{beenakker_colloquium:_2008} a clear
understanding of the effect of disorder in graphene Josephson junction is imperative.
In this paper we work directly at the tight-binding level and consider three
different types of
disorder: vacancies, ripples and impurity scatterers. In all of these cases we
find that disorder affects the Andreev bound states that are formed in the junction. For instance
vacancies induce a
zero-energy mode which destroys the Andreev gap when the concentration of
vacancies is increased. Similarly short-range scatterers and strong pseudo-magnetic fields will
broaden and subsequently destroy the Andreev peaks. We calculate the Josephson current induced by
the phase difference of the superconducting order parameter of the two contacts, and provide a
qualitative picture on the effect of various types of disorder.

This paper is organized as follows: In Sec. II we introduce our model and
the numerical approach
used. Results are organized according to the type of disorder considered. For
instance, results
concerning vacancies are presented and discussed in Sec. III, while the same is
done for the
cases regarding ripples and impurity scatterers in Sec. IV and V respectively.
Finally, we summarize our findings in Sec. VI.

\section{Model}
\begin{figure}
\includegraphics[width=.99\columnwidth,angle=0]{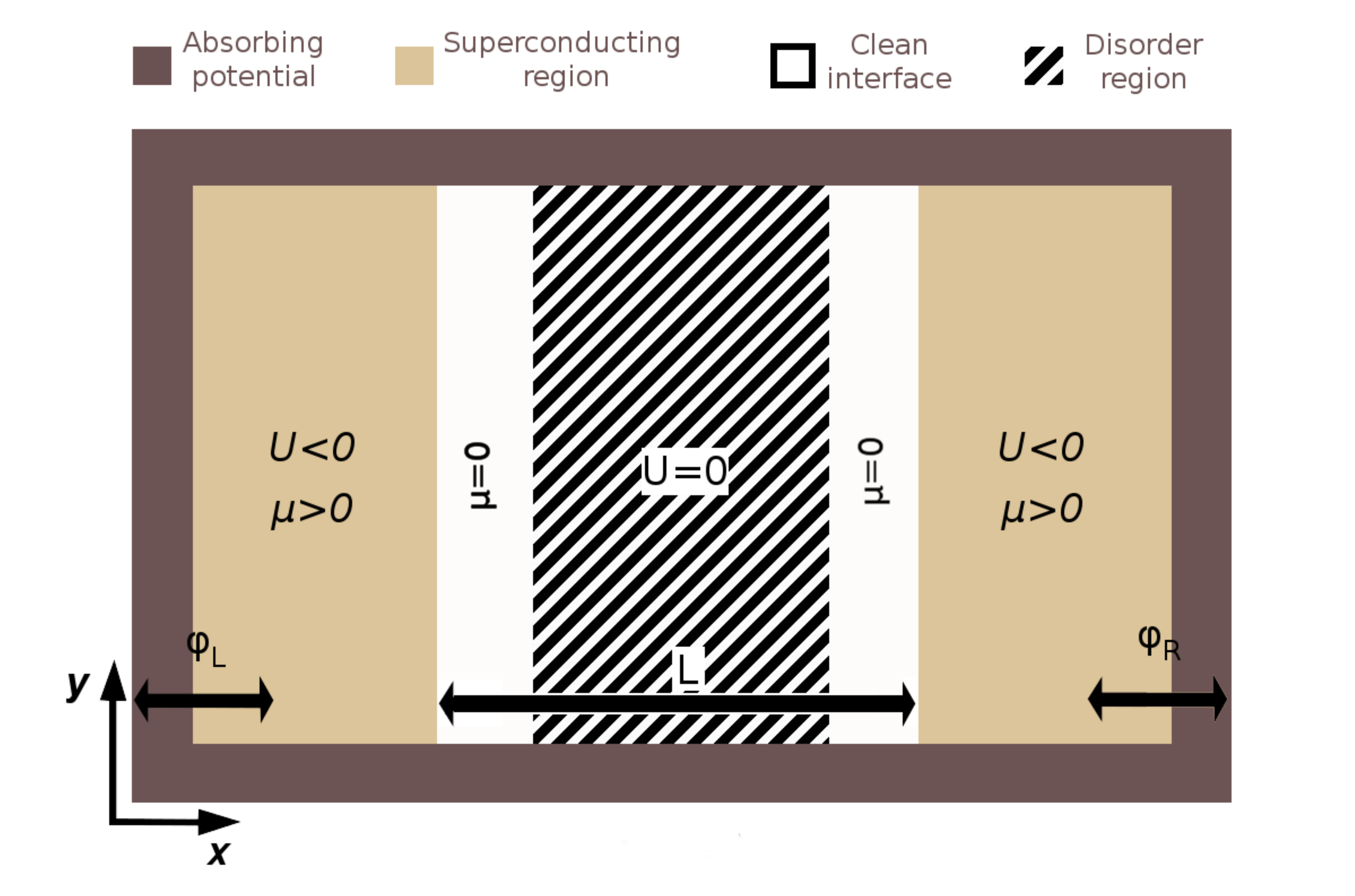}
\caption{ \label{layout} (Color online) Layout of graphene Josephson junction considered in this
work. Disordered and superconducting regions are
separated by clean interface strips where Andreev reflection takes place.
}
\end{figure}

A graphene Josephson junction is modeled according to the layout depicted in
Fig.~\ref{layout}.
Following closely the recipe implemented in previous
works~\cite{black-schaffer_self-consistent_2008,munoz_tight-binding_2012}, we
model the influence
of the right and left superconducting contacts by assuming an on-site attractive
pairing potential
$U<0$ and high doping, $\mu>0$. In this way, we introduce a
\textit{s}-wave superconducting state over the outermost regions in the graphene
sheet separated by a distance $L$. The width of the junction is considered to be much larger than
the
junction length, $W\gg
L$. We consider here an impurity-free S/N interface and set a high Fermi level
mismatch between
the superconducting and the normal parts of the junction where the paring potential
is set to zero $U=0$
(see Fig.~\ref{layout}). The large Fermi level mismatch  between the highly doped
contact region and the
undoped interface may suppress Andreev reflection from non-relativistic
electrons~\citep{akhmerov_pseudodiffusive_2007}.
We allow disorder only over the middle region of the junction, away from the
interfaces, in order to guarantee that the leakage of the Coopers pairs is
homogeneous along the clean interface strips.
In addition, an absorbing region is imposed at the borders of the junctions in order to
eliminate reflections coming from the boundaries and manifesting as finite size effects.

Finally,  a  dc-Josephson current is induced by fixing a phase difference
$\Delta\phi=\phi_R-\phi_L$ between the outermost parts of the contact regions, as shown in
Fig.~\ref{layout}. The calculation of the supercurrent is performed once the
amplitude and the phase of the order parameter is relaxed over the
superconducting region and convergence is achieved.
\subsection{The Chebyshev-Bogoliubov-de Gennes method.}
The Andreev scattering process \cite{andreev_thermal_1964} in graphene is described
within the Bogoliubov-de Gennes formalism by using  the  Nambu Gor'kov Green
functions \cite{nambu_quasi-particles_1960}:
\begin{eqnarray}
\label{nambu_gorkov_green_fn}
\gn_{ij} = -i\hbar\mat{\la Tc_{i\down}c^{\dagger}_{j\up}\ra}{\la
Tc_{i\down}c_{j\up}\ra}{\la Tc^{\dagger}_{i\down}c^{\dagger}_{j\up}\ra}{\la
Tc_{i\down}c^{\dagger}_{j\down}\ra},
\end{eqnarray}
where off-diagonal elements describe the amplitude of the superconducting order parameter coupling
electrons and holes. Other physical quantities, such as the local
density of states (LDOS), can be calculated from the Gor'kov Green's function
corresponding to the diagonal elements of the matrix in Eq.~(\ref{nambu_gorkov_green_fn}), which
is given by the expectation value of the inverse of the Hamiltonian:
$\gn=[\omega+\iu\eta-\ham]^{-1}$.

The elements of the matrix (\ref{nambu_gorkov_green_fn}) are calculated within the
approximation of the Gor'kov Green's functions by implementing the
Chebyshev-Bogoliubov-de Gennes method \cite{covaci_efficient_2010,
covaci_superconducting_2011, munoz_tight-binding_2012}:
\begin{eqnarray}
\label{cheb_expansion}
\bar{G}_{ij}^{1\alpha}(\tilde{\omega})=\frac{-2i}{\sqrt{1-\tilde{\omega}^2}}
\left[\sum_{n=0}^N
a_{ij}^{1\alpha}(n)e^{-in\arccos(\tilde{\omega})}\right],
\end{eqnarray}
where the expansion of the normal ($\alpha$=1) and anomalous
($\alpha$=2) coefficients corresponding to the diagonal and off-diagonal components of the
Green's function [\ref{nambu_gorkov_green_fn}] are define accordingly as:
\begin{eqnarray}
\label{normal}
a^{11}_{ij}(n) = \la c_{i\up}\left|T_n(\ham)\right|c_{j\up}^{\dagger}\ra, \\
\label{anomalous}
a^{12}_{ij}(n) = \la
c_{i\down}^{\dagger}\left|T_n(\ham)\right|c_{m\up}^{\dagger}\ra^{\ast}
\end{eqnarray}
\\
where $T_n(x)=\cos[n \arccos(x)]$ is the Chebyshev polynomial of the first kind of order $n$,
which  satisfies
the following recurrence relation: $T_{n}(x)=2xT_{n-1}(x)-T_{n-2}(x)$.

In the calculation of these moments it is necessary to normalize the
Hamiltonian matrix $\ham$ and its eigenvalues $\omega$, according
to $\tilde{\ham}=(\ham-\eins b)/a$ and $\tilde{\omega}=(\omega- b)/a$,
respectively, where the scale factors are $a=(E_{max}-E_{min})/(2-\eta)$ and
$b=(E_{max}+E_{min})/2$, with $\eta>0$ being a small number. Here, the extremal values
$E_{max}$ and $E_{min}$ bound the energy spectrum of the BdG Hamiltonian,
which in a real-space tight-binding formulation at the mean-field level can been written following
the
Nambu notation
(\ref{nambu_gorkov_green_fn}):
\begin{eqnarray}
\ham=\sum_{ \substack{<i,j>} } \vech{c_{i\up}^{\dagger}}{c_{j\down}}
\hat{\ham}_{ij}
 \vecv{c_{j\up}}{c_{j\down}^{\dagger}}
\label{bdg}
\end{eqnarray}
where the matrix $\ham$ can be seen as arrangements of matrix blocks as follows:
\begin{eqnarray}
\hat{\ham}_{ij}=\mat{\epsilon_i-\mu}{\Delta_i}{\Delta^{\ast}_i}{\mu-\epsilon_i}
\delta_{ij}+\mat{-t_{ij}}{0}{0}{t^{\ast}_{ij}}(1-\delta_{ij}),
\label{ham}
\end{eqnarray}
where $\epsilon_i$ denotes the on-site potential of the carbon atoms while $\mu$
 is the chemical potential which pins the Fermi energy. Nearest-neighbor
$p_{z}$-orbitals $A_i$ and $B_j$ are coupled through the hopping
parameter $t_{ij}$ which is known to be $t_{ij}$=$\gamma_0\approx$2.7eV for
pristine graphene, where the minimum distance between carbon atoms is assumed to
be $a_0\approx$1.42\AA.
Inhomogeneous superconductivity is taken into account through the on-site
s-wave order parameter $\Delta_i=U_i\la c_{i\up}c_{i\down}\ra$ where $U_i$
corresponds to the strength of the pairing potential and the complex correlation function
$\la c_{i\up}c_{i\down}\ra$ is derived from the Gor'kov Green function
[\ref{cheb_expansion}] according to the following equation:
\begin{eqnarray}
\la c_{i\up}c_{i\down}\ra = \frac{i}{2\pi}\int
\gn_{ii}^{12}(\omega)\left[1-2f(\omega)\right]d\omega,
\end{eqnarray}
where $f(\omega)$ is the Fermi distribution function. Another physically
relevant quantity is the local density of states, which is obtained from the Green
function (\ref{cheb_expansion}) through the following formula:
\begin{eqnarray}
\label{ldos}
N^i(\omega)=-\frac{2}{\pi}\text{Im}{G}_{ii}^{11}(\omega).
\end{eqnarray}
\subsection{Calculation of the moments}

Once the Hamiltonian has been normalized, the expectation values in
Eq.~(\ref{normal}) and (\ref{anomalous}), which define the expansion coefficients
$a_{ij}^{1\alpha}(n)=\la \alpha|\nu_n\ra$, where $\la\alpha|$ are the vectors
$\la 1|=\la c_{i\up}|$ and $\la 2|=\la c_{i\down}^{\dagger}|$, can be
straightforwardly obtained through an iterative procedure involving a successive
application of the Hamiltonian on the iterative vectors:
\begin{eqnarray}
\label{recur}
a(n)=\la
\alpha|\nu_n\ra=\la\alpha|\left(2\ham|\nu_{n-1}\ra-|\nu_{n-2}\ra\right).
\end{eqnarray}
with the initial conditions $|\nu_0\ra=|c_{j\up}\ra$ and
$|\nu_1\ra=\ham|\nu_0\ra$.
We refer the reader to Ref.~\onlinecite{covaci_efficient_2010} for more
details about the method.

In order to obtain the average DOS in the disordered region, we use a more suitable approach to
calculate the moments~(\ref{normal}). It was shown that the average DOS can be
expanded in terms of Chebyshev polynomials and the coefficients of order $n$ can be expressed as the
trace
of the polynomials of order $n$ of the Hamiltonian matrix. Instead of performing the full trace,
i.e. averaging over the LDOS, we perform a stochastic evaluation of the trace of the Hamiltonian as
follows~\cite{weise_kernel_2006}:
\begin{eqnarray}
\label{trace}
a(n)=\mathrm{Tr}[T_n(\ham))]\approx\frac{1}{R}\sum_{r=0}^{R-1}\la
r|T_n(\ham)|r\ra.
\end{eqnarray}
where the summation is carried out over $R$ random vectors $|r\ra$, which in an
arbitrary basis are defined as: $|r\ra = \xi_{ri}|c_{i\up}^{\dagger}\ra$
with random coefficients $\xi_{ri}$ having a normal random
distribution over the interval $[-1,1]$.  The  statistical average over
the $R$ vectors of the expectation value approximates the trace in Eq.~(\ref{trace}).
Here the number of vectors $R$ required to perform (\ref{trace}) is much lower than
the order $M$ of the Hamiltonian~(\ref{bdg})
($R<M$)~\cite{iitaka_random_2004} and the number of atoms in the disordered region.

The average DOS is calculated by using Eq.~(\ref{ldos}) with the moments
obtained using Eq.~(\ref{trace}). The number of random
vectors, $R$, is taken to be large enough such that the DOS converges. We typically
use $R\sim200$.

Since most of the computational effort comes from the sparse Hamiltonian
matrix multiplications with iterative vectors, the performance is dramatically
increased by implementing a parallel algorithm on graphical processing units
(GPUs) using CUDA.
\subsection{Complex absorbing potential}
In order to mimic effectively an infinite region we introduce an absorbing
potential operator $\hat{\Gamma}$ at the boundary by following the recipe
of Ref.~[\onlinecite{mandelshtam_simple_1995}].
A proper choice of $\Gamma$ requires its imaginary part to be negative
(Im$\hat{\Gamma}\leq 0$) with an arbitrary choice for its real part, which we set to
zero (Re$\hat{\Gamma}=0$) for convenience. It is expected that a well-behaved
absorbing potential will eliminate reflection effects at the boundary.
It can be shown that this absorbing boundary condition could be easily incorporated in the
Chebyshev expansion of the Green's function by considering an
imaginary damping factor $\hat{\gamma}$, which redefines the recursion
formula~(\ref{recur}) as follows:
\begin{eqnarray}
\label{cap}
|\nu_{n}\ra = e^{-\hat{\gamma}}\left(2\ham|\nu_{n-1}\ra -
e^{-\hat{\gamma}}|\nu_{n-2}\ra\right),
\end{eqnarray}
with the initial conditions $|\nu_0\ra=|c_{i\up}^{\dagger}\ra$ and
$|\nu_1\ra=e^{\hat{-\gamma}}|\nu_0\ra$. Our calculations for pristine graphene
(not presented here) show that we can remove all the finite size
interference peaks in the LDOS, appearing due to scattering from the boundaries, without
implementing
periodic boundary conditions or considering large lattice sizes.
\section{Results}
\subsection{Vacancies}
Within the tight-binding formalism considered here, a vacancy at atomic site $i$ is modeled by
setting the on-site energy larger that any energy scale present in the pristine normal state in
graphene, $\epsilon_i \gg 3t$.
In addition, the corresponding hopping parameters that connect the $i$-site to its neighbors are
set to zero, $t_{ij}=0$. Our numerical procedure considers a finite concentration of
vacancies, defined through the ratio $x_{vac}=N_{vac}/N$, where $N$ is the total number of atomic
sites in the junction.\\
\begin{figure}[ttt]
\includegraphics[width=1.05\columnwidth]{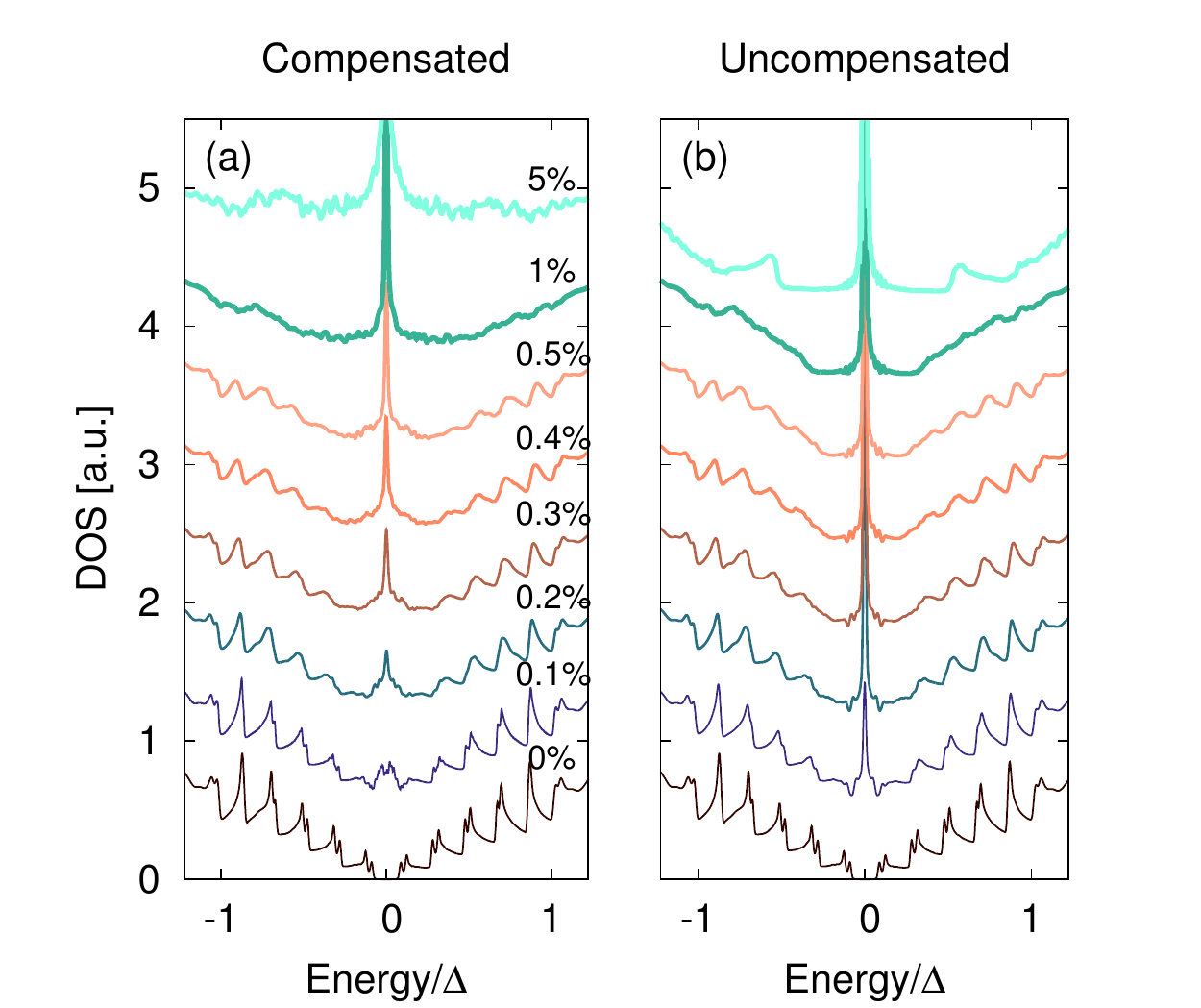}
\caption{ \label{dos:vac} (Color online) DOS for a graphene Josephson junction considering different
concentration of vacancies $x_{vac}=N/N_{vac}=0,0.1,0.2,0.3,0.4,0.5,1$ and $5\%$ from bottom to top.
Two cases are considered, according to the compensation in the distribution between the two
sublattices:  (a) completely compensated where vacancies are distributed equally in the two
sublattices
and (b) completely uncompensated where vacancies are randomly present only in one of the
sublattices.
}
\end{figure}
It has been shown that divergences in the DOS are present around the Dirac point when a finite
concentration
of vacancies are induced in graphene~\cite{pereira_disorder_2006}. Here, we calculate the DOS of
the graphene Josephson junction while introducing different concentrations of vacancies over the
disordered region depicted in Fig.~\ref{layout}. The DOS is averaged over different realizations
for the configuration of vacancies in order to include all possible scattering and interference
processes between the electron/hole wave and the scatterers induced by the missing atoms. In
Fig.~\ref{dos:vac} we shown the different Andreev peaks that are present in our clean Josephson
junction
of graphene (bottom curve). The electron and hole wave are scattered by the N/S
interface and interfere, giving rise to the well known Andreev peaks. Note that because the
chemical potential in the junction is at the Dirac point, the Andreev peaks show the specific
energy dependence in the graphene junctions, i.e. traveling modes with a gapped spectrum versus
bound
states with an ungapped spectrum as observed in conventional S/N/S junctions. As we can see clearly
in Fig~\ref{dos:vac}, as the concentration of impurities increases, the Andreev peaks are
progressively suppressed. Notice that the lowest energy peaks are the first to disappear as
the scattering probability for long paths between the interfaces becomes higher, affecting lower
energy states, when compared to short paths, which contribute to higher subgap energy states.
As the concentration is increased above $x>0.5\%$, traces of the Andreev reflection processes
on the DOS vanish.

\begin{figure}[ttt]
	\includegraphics[width=.99\columnwidth]{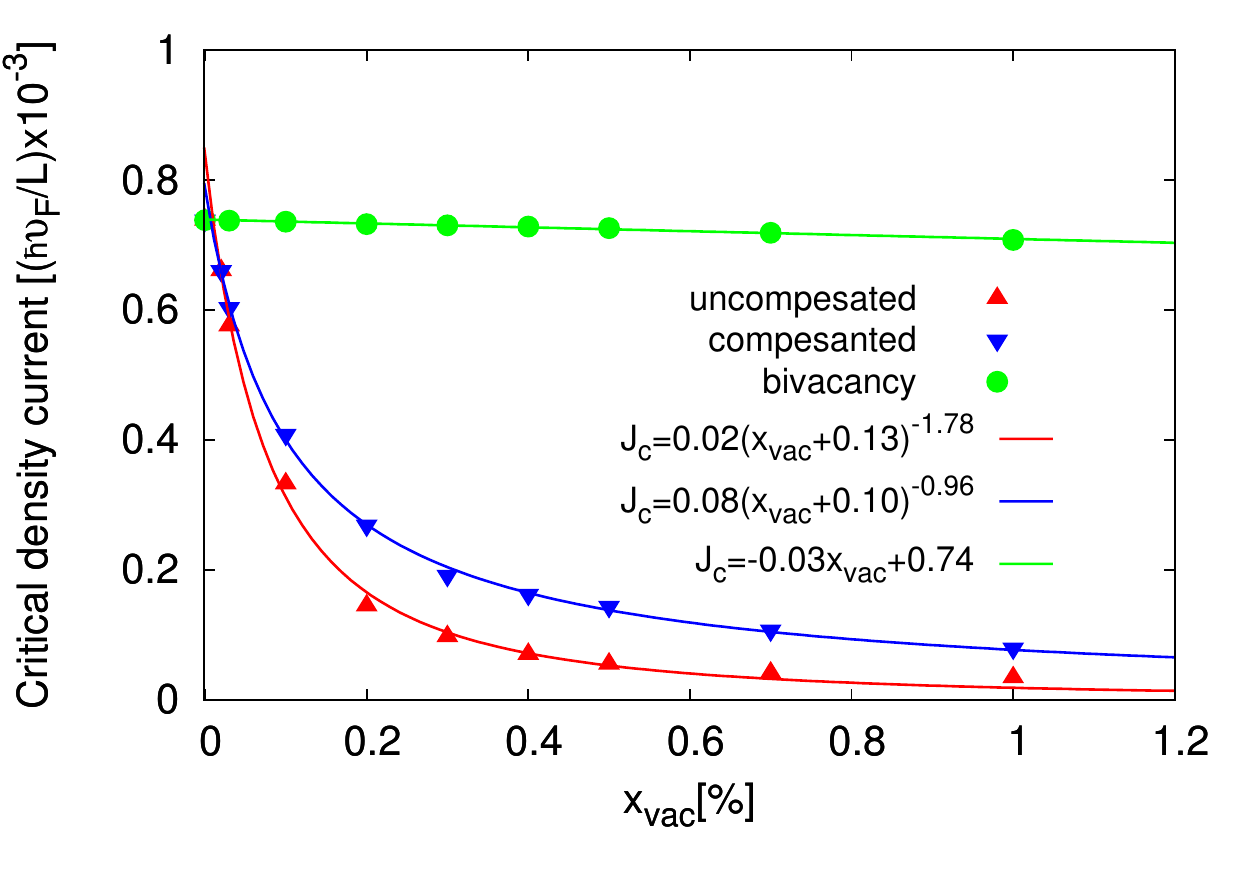}
\caption{(Color online) Average critical current density in a graphene Josephson junction for
different concentrations of vacancies, diluted randomly over both sublattices (compensated), or over
a single sublattice (uncompesated), or over two-coupled sublattices (bivacancy). The continuous
lines correspond to a least squares fit to $f(x)=a(x+b)^{c}$ and $g(x)=ax+b$.
}\label{cur:vac}
\end{figure}

Selective dilution of the vacancies in the different
sublattice sites has been shown to induce different zero modes in the
DOS~\cite{pereira_disorder_2006,Cresti2013}. For instance, complete
dilution of the vacancies in one of the
sublattices, or complete uncompensated dilution, leads to the opening of a gap around the zero mode
at the Dirac point and whose magnitude is proportional to the vacancy concentration. On the
opposite, complete compensated dilution brings an increase of the spectral weight for
energies surrounding the Dirac point. We have investigated both cases and found slight
differences in the DOS around the Fermi level, which are more remarkable for larger vacancy
concentrations, $x>0.5\%$ (see Fig.~\ref{dos:vac}). In order to further clarify the contrast
between the compensated and uncompensated cases we show in Fig.~\ref{cur:vac} the
average critical current density across the junction as a function of vacancy concentration.
Note that the current is averaged over the junction width and over impurity configurations.
As we can observe both cases lead to different power-law suppression of the the critical
current. For instance, the suppression of the current goes according to $J_c\sim x^{-1.8}$ for
uncompensated dilutions of the vacancy configurations. This agrees with the fact that a gap of
energy scale $E^2\sim x_{vac}$ is induced for complete uncompensated dilution of vacancies in
graphene~\cite{pereira_disorder_2006}. On the other hand, vacancies diluted in both sublattices
still show a strong suppression of the supercurrent but weaker than the uncompensated case. As an
interesting fact, we note that by placing these vacancies in pairs of bounded atomic sites
(bivacancies), we can observed that the suppression is much weaker than in previous cases.
The slow linear suppression of the supercurrent in the presence of bivacancies is due to the absence
of inter-valley scattering, as was already reported previously for this type of
atomic-scale defects~\cite{park_formation_2011}.
\subsection{Ripples: Gaussian bumps}
\begin{figure}[hhh]
\centering
	\includegraphics[width=0.8\columnwidth,angle=0]{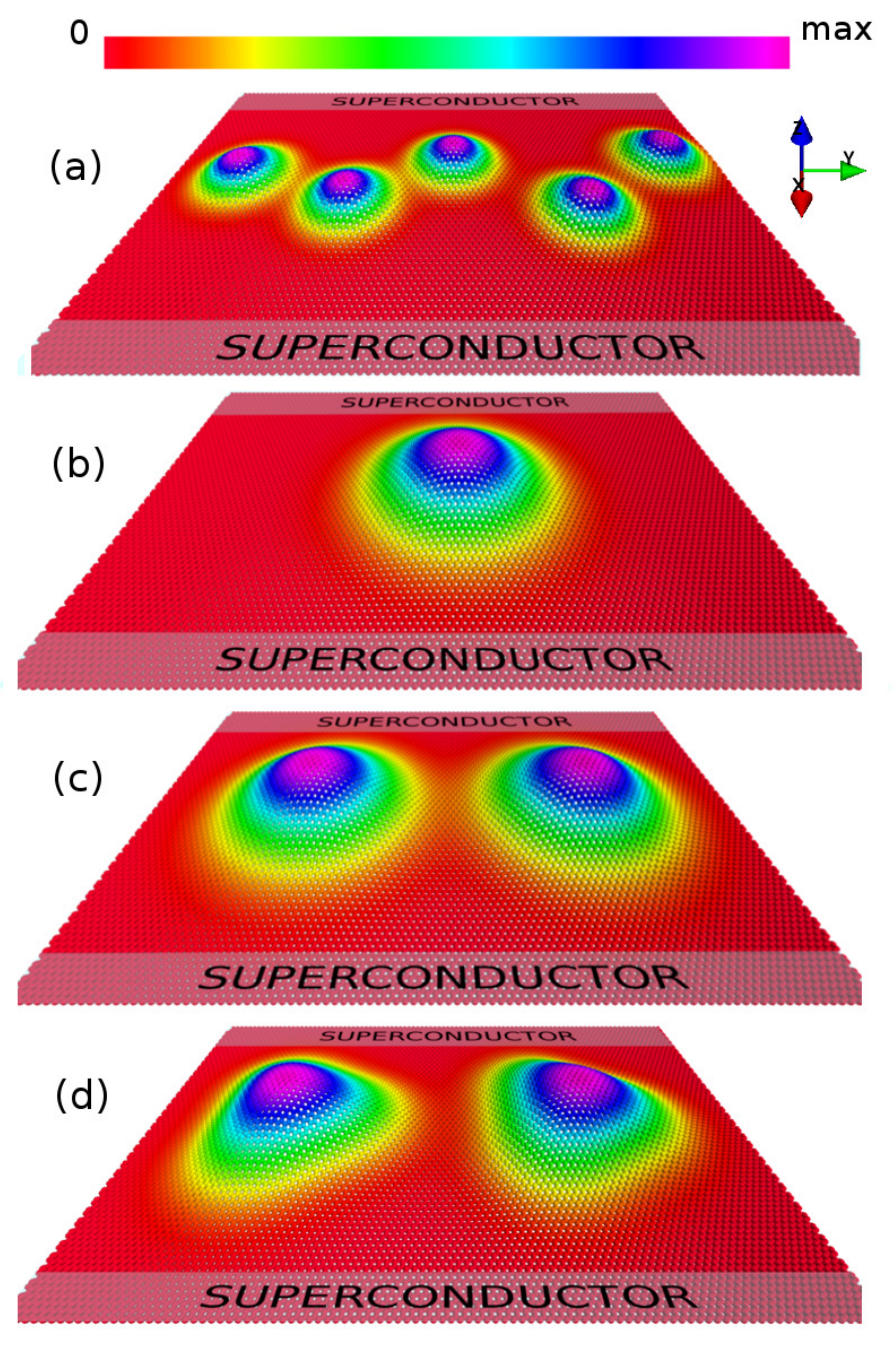}
\caption{  \label{bumps} (Color online) Strained graphene Josephson junction with different
configurations of Gaussian bumps. Panels (a)-(c) show isolated Gaussian bumps while (d) corresponds
to an arrangement of four Gaussian bumps in a triangular configuration.
}
\end{figure}

\begin{figure}[ttt]
\vspace{-0.2cm}
	\includegraphics[width=1.1\columnwidth]{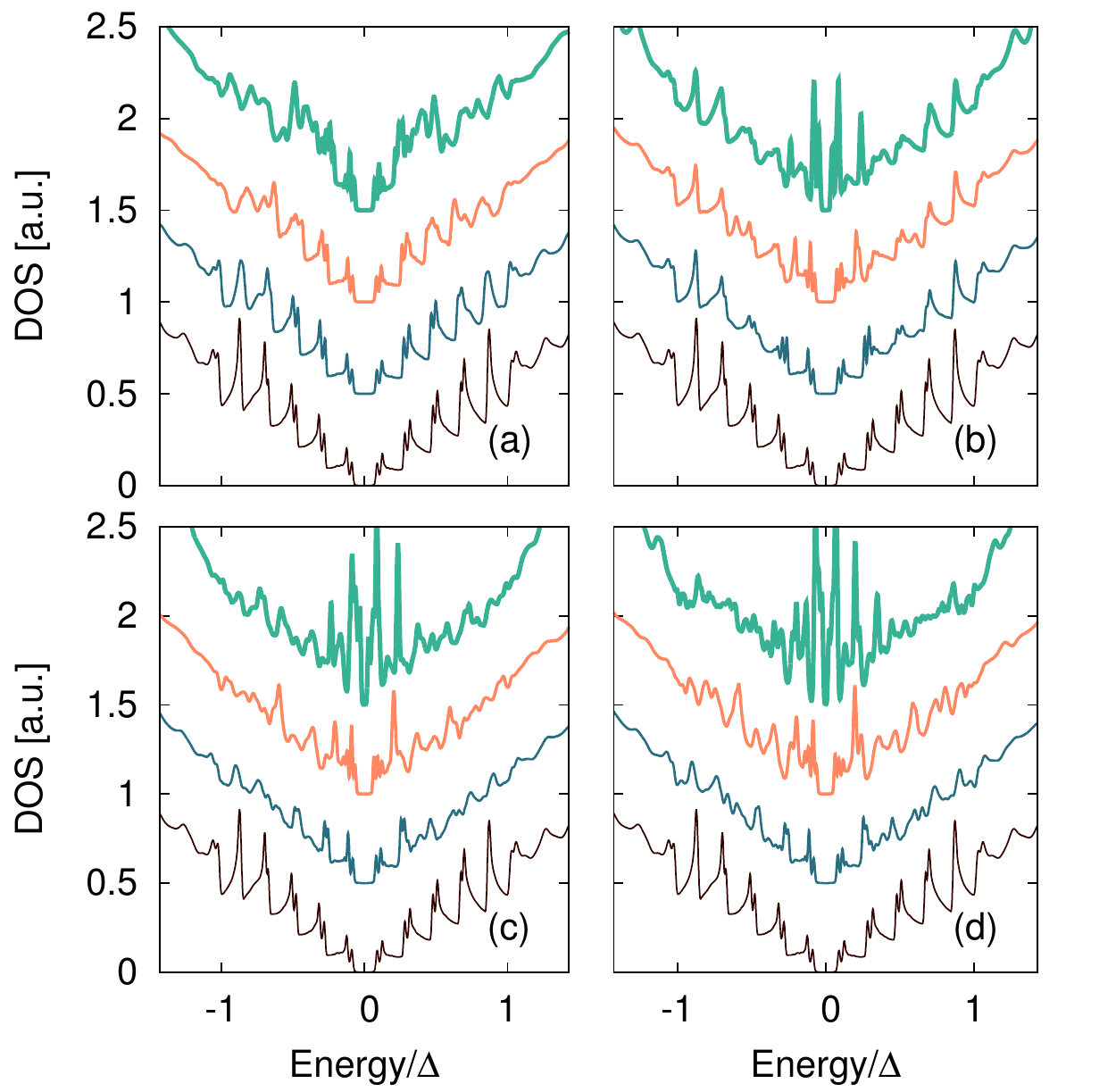}
\caption{\label{dos::ripp}(Color online) DOS for a graphene Josephson junction for the
different strain configurations shown in Fig.~\ref{bumps}. Panels from left to right
correspond to cases from top to bottom shown in Fig.~\ref{bumps}, respectively. Areas delimited by
subsequent curves have been filled in order to present a better contrast between the Andreev
peaks for deformations with the maximum strain increasing from 0\% to
5\%, 10\% and 20\% (bottom to top).
}
\end{figure}

We next analyze the effects of inhomogeneous strain over the disordered area in the graphene
Josephson junction. For this purpose, we model the ripples in graphene as
smooth bumps where the out of plane deformation is described by a Gaussian function. It has been
theoretically shown that a six-fold spatially symmetric pseudo-magnetic fields, with
alternating sign, emerges from this strain
configuration~\cite{moldovan_electronic_2013,Wakker2011,Klimov2012}. \\
The Gaussian deformation is introduced in the tight-binding description of Eq.~(\ref{ham}) by the
strained hopping parameter:
\begin{eqnarray}
t_{ij}=\gamma_0\exp^{-3.37(\frac{l_{ij}}{a_0}-1)}
\end{eqnarray}
where $\gamma_0=2.7$eV and $a_0=1.42\AA$ are the unstrained hopping and lattice parameter,
respectively, while $l_{ij}$ is the strained distance between nearest-neighbors $i$ and $j$. The
corresponding out of plane deformation is given by a Gaussian function as follows:
\begin{eqnarray}
Z(R_{ij})=Z_0\exp^{-|R_{ij}|^2/2\varepsilon^2}.
\end{eqnarray}
where $R_{ij}=r_{ij}-R_0$ is the in-plane atomic position with respect to the center of the
Gaussian, $R_0$.
It is important to mention that the Gaussian width parameter $\varepsilon$ is constrained here, such
that $Z(R_{ij})\approx 0$ in the clean interface regions (see Fig.~\ref{layout}).
Once the width is fixed, the height parameter $Z_0$ is adjusted according to a desired maximal
strain. Since we known from the continuum model how the strength of the pseudomagnetic field
depends on the parameters of the Gaussian, we considered different configurations for the
size and the number of Gaussian bumps inside the junction (see Fig.~\ref{bumps}). As a
particular case, we have included in Fig.~\ref{bumps} an arrangement of triangular bumps made from a
superposition of four Gaussians in a triangular configuration, where three are centered in
equidistant vertices while the last is placed a distance $d$ from the vertices in the middle of
the triangle. This particular strain has been inspired by a previous theoretical study where a
nonuniform deformation is engineered by depositing graphene on a substrate decorated with
nanopillars set in a triangular
configuration~\cite{neek-amal_nanoengineered_2012}. The corresponding pseudomagnetic fields emerging
from these deformation exhibit a non-trivial symmetry, consisting of larger regions with an
almost constant pseudo-magnetic field, when compared to the sixfold symmetric fields generated by
isolated Gaussian bumps.
In order to investigate the effect of ripples on Andreev scattering in the junction we calculated
the average DOS for the different configurations of Gaussian bumps shown in Fig.~\ref{bumps}.
These results are shown in Fig.~\ref{dos::ripp}(a)-(d) for different values of the maximum strain:
0\%, 5\%, 10\% and 20\%. The Andreev states seen in Fig.~\ref{dos::ripp}(a) start
to be affected by the presence of
the Gaussian bumps even for the lowest values of the strain. Particularly, high-energy Andreev peaks
are
mainly suppressed for the case of high density of Gaussian bumps depicted in Fig.~\ref{bumps}(a).
This suggests that short paths are influenced more by the pseudomagnetic field than the long paths,
which contribute to the low energy spectrum.

Alternatively, the configuration with a single Gaussian bump, shown in Fig.~\ref{bumps}(b), for
which the DOS is presented in Fig.~\ref{dos::ripp}(b) shows that high-energy Andreev peaks remain
conserved despite the fact that the pseudomagnetic field is supposed to be stronger over a wider
region. At low energies, sharp peaks appear, and the quasi-particle gap starts to close.

\begin{figure}[hhh]
\vspace{-0.2cm}
	\includegraphics[width=1.\columnwidth]{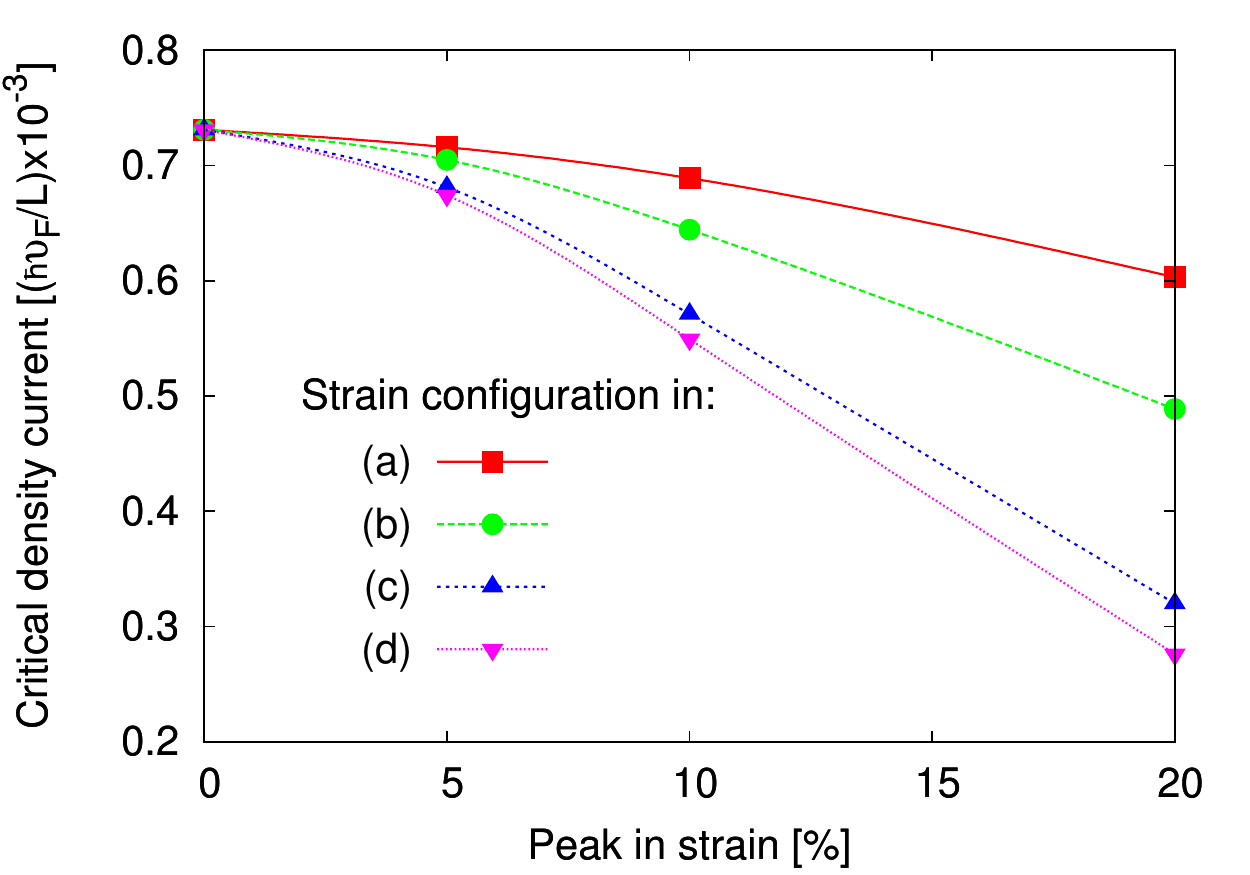}
\caption{\label{cur::ripp}(Color online) Average critical current density in strained
Josephson
junctions as a function of maximal strain for the strain configurations depicted in
Figs.~\ref{bumps}(a)-(d).
}
\end{figure}

The next two cases, shown in Figs.~\ref{bumps}(c) and (d) and Figs.~\ref{dos::ripp}(c) and (d),
reveal an interplay between the size of the Gaussian bump and the strength of the pesudomagnetic
field.
As seen previously, the high energy peaks are the first to be affected as the strain is increased.
At low energies, Landau level-like peaks appear and the gap observed for the clean system closes.
Because of the oscillating pseudo-magnetic field, a combination of snake-like
states and pseudo-Landau levels, appear where the field vanishes or is maximal. The existence of
extended regions with large pseudo-magnetic fields, will act as a pseudo-magnetic barrier for the
electron or hole quasiparticles propagating in the junction. As a consequence, although the
time-reversal symmetry is not broken, it is expected that the supercurrent is suppressed and will
flow only as edge states near the boundaries~\cite{covaci_superconducting_2011}.
%
%
We next investigate the Josephson current flowing through the strained junctions in
Fig.~\ref{cur::ripp}, where we present the average critical
current density as a function of the maximum strain applied in each configuration.  We find a
suppression of the critical current as the strain is increased. In particular, this suppression is
stronger for the configurations shown in Figs.~\ref{bumps}(c) and (d) when compared with the single
bump case, Fig.~\ref{bumps}(b), and the many smaller bumps, Fig.~\ref{bumps}(a). As suggested
previously, these results can be explained by noticing that the current is more effectively
suppressed in the cases in which the pseudo-magnetic field is larger over a more extended
area, thus providing a better magnetic barrier. Although the size of the bump in case (d) is the
same as the one in case (c), the current is further suppressed because the triangular bump induces
 larger regions with pseudo-magnetic fields with the same sign, although the average field is zero.
\subsection{Charged impurities}
Finally, we investigate the effect of scattering due to the presence of impurities on the diffusion
of Andreev pair in graphene Josephson junctions. It is well-known that the presence of these types
of impurities may induce local charged puddles in graphene, which can be emulated through spatial
fluctuations of the Fermi energy around the Dirac point. These fluctuations are modeled here by
means of a random superposition of $N_{imp}$  potentials with a Gaussian-like spatial dependence.
Thus, we assume that the proximity of a single charged impurity is reflected in the parameters of
a Gaussian potential such that on-site energies of carbon atoms in the disordered region are given
as follows~\cite{rycerz_anomalously_2007}:
\begin{eqnarray}
 \label{imp_pot}
 \epsilon_i = \sum_{j=1}^{N_{imp}}V_j \exp\left(-\frac{|r_i-R_j|^2}{2\varepsilon^2}\right)
\end{eqnarray}
where $V_j$ and $\varepsilon$ correspond to the amplitude and range of a
single Gaussian potential centered at the atomic position
 $R_j$, respectively. These on-site potentials as described by Eq.~(\ref{imp_pot}) are introduced in
our formalism described by the Hamiltonian (\ref{ham}) but we constrain their scope to
$\varepsilon < L, W$ such that they vanish in the interface strips where the Fermi level is pinned
at the Dirac point.

In order to characterize the effect of this sort of disorder we follow the recipe proposed in
previous works ~\cite{rycerz_anomalously_2007,lewenkopf_numerical_2008} where the mean-free path is
considered as being inversely proportional to the following parameter:
\begin{eqnarray}
 \kappa_0 \propto \left(\frac{V_i}{t}\right)^2 x_{imp}\kappa^2
\end{eqnarray}
where $V_i/t$ is  amplitude of the random potential in units of the hopping parameter while
 the ratio of the number of impurities and the total number of atoms in the junction corresponds to
 the concentration of impurities $x_{imp}=N_{imp}/N$. The averaged charge density per impurity atom
is described by the factor $\kappa$ according to:
\begin{eqnarray}
\label{charge_den}
\kappa=\frac{1}{N_{imp}}\sum_{j}^{N_{imp}}\sum_{i}^{N}
\exp\left(-\frac{|x_i-x_j|^2}{2\varepsilon^2}\right)
\end{eqnarray}
For practical purposes, we consider the same amplitude for all potentials in Eq.~(\ref{imp_pot}),
\textit{i.e.} $|V_j|=V$ for all $j$. In addition, for a given value of the Gaussian width
$\varepsilon$, we fixed the maximum of the Gaussian, $V$, such that the density of charges obtained
from Eq.~(\ref{charge_den}) is the same for the different values of $\varepsilon$ considered here.
This allows us to make a more clear discussion about the effect of the Gaussian potential, mainly of
its
height and width parameters,  as long as the total charge density is kept fixed under a constant
concentration of vacancies. We consider two separate cases. First, in the presence of electron and
hole puddles, the total charge density is zero, meaning that the number of electron and hole-doped
Gaussians is equal. We next investigate the presence of charged impurites of the same type, i.e.
electron-doped Gaussians, in which case the charge density will increase as the concentration of
impurities increases.

\begin{figure}[ttt]
	\includegraphics[width=1.1\columnwidth]{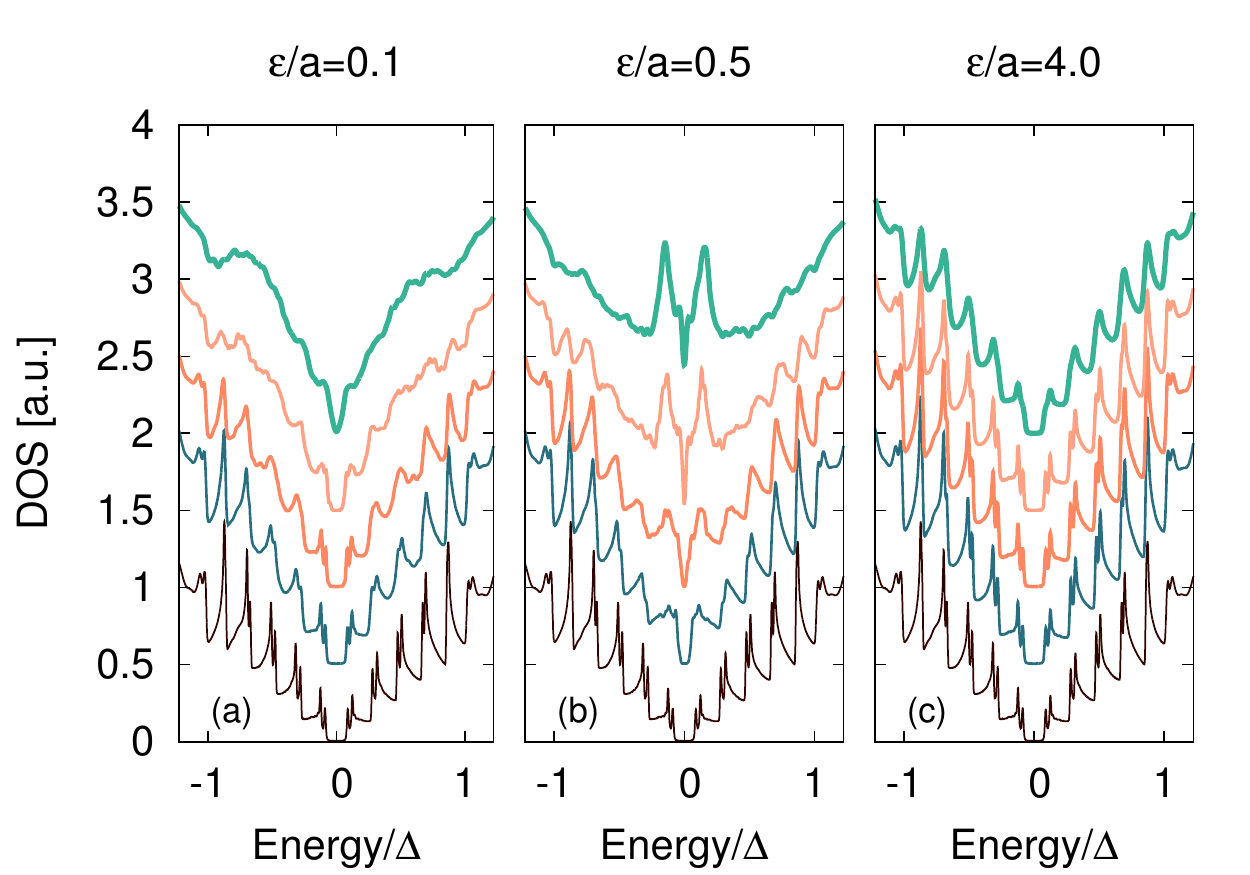}
\caption{\label{dos::pock}(Color online) Average DOS for disordered graphene Josephson junctions,
considering impurity scattering potentials with different ranges, $\varepsilon/a=$0.1,
0.5 and 4.0. The area between the curves has been filled in order to have a better contrast
for lines corresponding to different impurity concentrations. Different values of the impurity
concentration considered inside the panels are $x_{imp}=$0\%, 1\%, 2\%, 3\% and 5\%, from bottom to
top.
}
\end{figure}

We first present our results for the electron-hole charge puddles. In Figs.~\ref{dos::pock}(a)-(c)
we shown the average DOS in a graphene Josephson junction with doping inhomogeneity profiles given
by
Eq.(~\ref{imp_pot}) for three different cases according to the size of the Gaussian potential
induced by single impurities. As in the previous disorder cases, the DOS is averaged
over different realizations of the impurity configurations.
The most trivial case, where $\varepsilon/a=0.1\ll 1$, is depicted in
Fig.~\ref{dos::pock}(a). In this limit, which resembles the typical Anderson disorder model,
we can observe that the effect of increasing the concentration of impurities
 $x_{imp}$ leads to a suppression of the Andreev bound states. Despite that this on-site
 defect is similar to the case of vacancies (in the limit of large $V$) we can see that
when comparing with Fig.~\ref{dos:vac} the dispersion mechanism acts differently in the two cases,
as lower energy Andreev states are preserved even for the highest values of $x_{vac}$ in
 Fig.~\ref{dos::pock}(a).

Next we proceed to a larger $\varepsilon/a=0.5$, a value still smaller than the
lattice parameter. In this situation,  we see in Fig.~\ref{dos::pock}(b) that Andreev
bound states are affected much more strongly as inter-valley scattering is expected to become more
pronounced. It is clearly seen that the lower energy gap disappears in the first place
as the concentration of impurities is increased. This is similar to what was observed for
vacancies, with the difference that the quasiparticle gap does seem to be completely suppressed as
the impurity concentration increases.

\begin{figure}[ttt]
	\includegraphics[width=1.0\columnwidth]{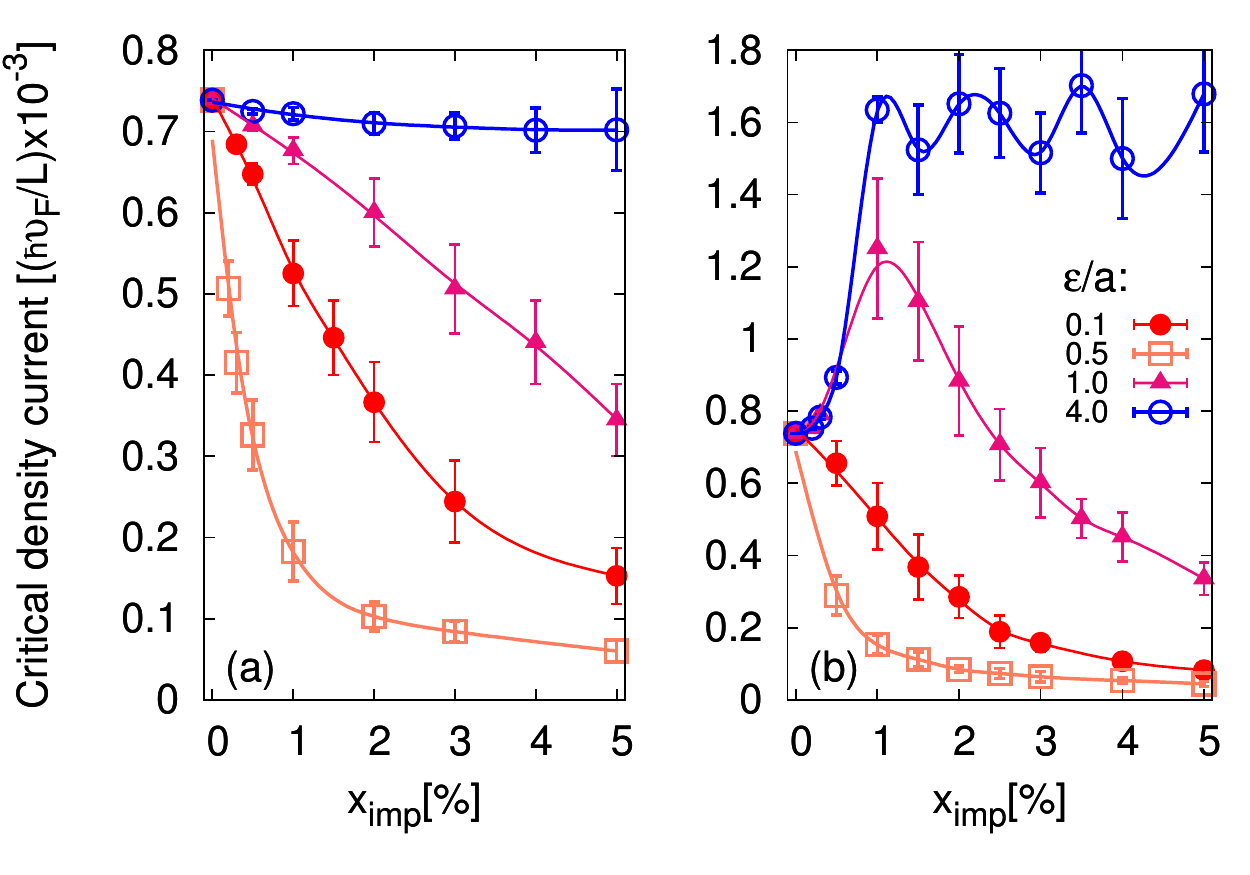}
\caption{  \label{curr::pock}(Color online) Average critical current density in a graphene
Josephson
junction as a function of the impurity concentration $x_{imp}$ for different widths of the  Gaussian
potential induced by isolated impurities. Equal number of electron and hole-like potentials are
considered in panel (a) while only electron-like potentials are assumed in (b) leading to a
inhomogeneous but finitely doped junction.
}
\end{figure}

Contrary to the previous cases, when $\varepsilon>a$, the influence of disorder on the Andreev
states is weak. As the potential profile becomes smoother on the scale of the lattice
parameter, inter-valley scattering is suppressed, thus having a weak influence on the average DOS in
the junction, as seen in Fig.~\ref{dos::pock}(c).
In order to verify the insights given by the change in the Andreev levels seen in the averaged DOS,
we plot in Fig.~\ref{curr::pock} the
average critical current density as a function of impurity concentration for different
values of $\varepsilon$. First, in panel Fig.~\ref{curr::pock}(a), the electron-hole puddles
scenario is considered. In this case the current is suppressed for all ranges of the potential
profiles, with a much stronger effect when $\varepsilon \leq a$, and a very weak effect when
$\varepsilon>a$. The case $\varepsilon/a=0.5$ deserves particular attention as the current becomes
strongly suppressed when the impurity concentration is increased, confirming the strong suppression
of the Andreev peaks observer in the averaged DOS. This effect was investigated experimentally
in graphene Josephson junctions in the long-junction limit~\cite{komatsu_superconducting_2012} and
presence of electron-hole puddles was given as an explanation for the strong suppression of the
supercurrent at charge neutrality. As shown here, we only observe a strong suppression when the
charge scatterers are short-range.

In addition to the charge puddles case, the effect of an impurity distribution with positive
charge, inducing an average finite doping in graphene, is also studied. The average critical current
density for this scenario is presented in Fig.~\ref{curr::pock}(b). Here the dependence of the
current is not monotonic as a function of $\varepsilon$. The overall tendency is for the
supercurrent to be enhanced since finite doping brings the Fermi level away from the Dirac
point. On the other hand, the presence of short-range scatterers will also generate inter-valley
scattering events, thus suppressing the current. Therefore we observe two separate regimes,
depending on whether $\varepsilon < a$ or $\varepsilon \geq a$. For $\varepsilon=0.1a$ and $0.5a$,
the current is suppressed even for low impurity concentration, signaling the fact that inter-valley
scattering is the main contribution, larger for $\varepsilon=0.5a$. If $\varepsilon \geq a$, at low
impurity concentration the supercurrent is increasing due to an increase in the average doping but
depending on the inter-valley scattering probability it will be eventually suppressed, more for
$\varepsilon = a$ than $\varepsilon=4a$.

\section{Conclusion}
In conclusion, by using a numerical tight-binding approach, we described various disorder scenarios
in graphene Josephson junctions near charge neutrality. We investigated both the disappearance of
the multiple Andreev reflection peaks in the junction and the suppression of the Josephson current.
We observed that the supercurrent is most strongly suppressed in the presence of vacancies or
resonant impurities, e.g. adsorbed hydrogen atoms. In this case, the presence of strong inter-valley
scattering destroys the interference of time reversed electron-hole pairs which undergo Andreev
reflections at the N/S interfaces. As a test, we show that when the vacancies come in pairs, and
thus the sub-lattice symmetry is not being broken, the supercurrent is very weakly suppressed.

Another scattering
mechanism is given by the presence of ripples. We show that although there should be no
inter-valley scattering in this case, Gaussian bumps will act as
pseudo-magnetic barriers, thus suppressing the supercurrent. The larger the regions with finite
pseudo-magnetic fields, the more efficient the scattering will be.

A third disorder scenario involves the presence of charged impurities, which are modeled as
variations of the local potential. We show that in the presence of electron-hole charge puddles,
the supercurrent is always suppressed, but the strongest effect is obtained when the range of the
potential disorder is very small, thus inducing significant inter-valley scattering. When the
impurities only dope with electrons, we observe an interplay between an enhancement of the current
due to the shift of the Fermi energy away from the Dirac point, and a suppression by short range
scatterers due to inter-valley scattering.
\begin{acknowledgments}
This work was supported by the Flemish Science Foundation (FWO-Vlaanderen) and the Methusalem
funding of the Flemish Government.
\end{acknowledgments}
%
%

\end{document}